\documentclass[11pt]{article}
\usepackage{amssymb}
\usepackage{amsmath}
\usepackage{graphicx}
\usepackage{bm}
\newcommand{\mathsym}[1]{{}}

\usepackage{graphicx}
\usepackage{rotating}
\usepackage{color}
\usepackage{cancel}
\newcommand{\bra}{\begin{array}}
\newcommand{\era}{\end{array}}
\newcommand{\beq}{\begin{equation}}
\newcommand{\eeq}{\end{equation}}
\newcommand{\beqar}{\begin{eqnarray}}
\newcommand{\eeqar}{\end{eqnarray}}

\def\BC{\bb C}
\def\_\BC{\bbi C}

\def\( {\left(}
   \def\) {\right)}
\def\[ {\left[}
\def\] {\right]}

\def\dag {{\dagger}}

\newtheorem{proposition}{Proposition}[section]
\newtheorem{theorem}{Theorem}[section]

\newtheorem{remark}{Remark}[section]

\begin{document}
\begin{titlepage}

\vspace{20pt}

\begin{center}

{\LARGE \bf ${\theta}(\hat{x},\hat{p})-$deformation  of the  harmonic oscillator in a $2D-$phase space\\
\medskip
 }
\vspace{15pt}

{\large M. N. Hounkonnou$^{\dag,1}$  D. Ousmane Samary$^{\ddag,1,2}$,\\ E. Balo\"itcha$^{*,1}$ and ,  S. Arjika$^{**,1}$} 

\vspace{15pt}

{\sl $1$-International Chair in Mathematical Physics and Applications\\ (ICMPA-UNESCO Chair), University of Abomey-Calavi,\\
072B.P.50, Cotonou, Rep. of Benin}\\

{\sl $2$-CNRS - Universit\'e Lyon 1, Institut Camille Jordan\\ (Bat. Jean Braconnier,  bd du 11 novembre 1918),\\
F-69622 Villeurbanne Cedex, France}\\

\vspace{5pt}
E-mails:  {\sl 
$^{\dag}$norbert.hounkonnou@cipma.uac.bj , \sl 
$^\ddag$ousmanesamarydine@yahoo.fr, \\
\sl $^*$ezinvi.baloitcha@cipma.uac.bj,  \sl $^{**}$rjksama2008@gmail.com}

\end{center}

\vspace{10pt}
\date{\today}
\begin{center}
[Ref-preprint]
CIPMA-MPA/014/2012
\end{center}
\begin{abstract}
This work addresses a ${\theta}(\hat{x},\hat{p})-$deformation  of the harmonic oscillator in a $2D-$phase space. Specifically,
 it concerns a quantum mechanics of the harmonic oscillator based on a noncanonical commutation relation depending
 on the phase space coordinates.
 A reformulation of this deformation is considered in terms of
a $q-$deformation allowing to easily deduce the energy spectrum of the induced deformed harmonic oscillator. 
Then, it is proved  that the deformed position and momentum operators admit a one-parameter family of self-adjoint extensions. 
These operators  engender new families of deformed Hermite polynomials generalizing usual $q-$ Hermite polynomials.
Relevant matrix elements are computed. Finally,  a $su(2)-$algebra representation 
of the considered
deformation is investigated and discussed.
\end{abstract}

\noindent  Pacs numbers: 
\\
\noindent  {\bf Key words}: 
Harmonic oscillator, energy spectrum, q-deformation, Hermite polynomials, matrix\\ elements,  $su(2)-$algebra

\setcounter{footnote}{0}

\end{titlepage}
\section{Introduction}
 Consider a $2D$ phase space $\mathcal{P}\subset\mathbb{R}^2.$ Coordinates of position and momentum are denoted by
$x$ and $p.$ Corresponding Hilbert space quantum operators $\hat{x}$ and $\hat{p}$ satisfy the following commutation relation
\begin{eqnarray}\label{sos0}
[\hat{x},\hat{p}]=\hat{x}\hat{p}-\hat{p}\hat{x}=i\theta(\hat{x},\hat{p})
%\equiv i(1+\alpha\hat{x}^2+\beta\hat{p}^2)
\end{eqnarray}
where
$\theta$ is the deformation parameter
 encoding the noncommutativity of the phase space:
$\theta(x,p)=1+\alpha x^2+\beta p^2,\,\,\alpha,\beta\in\mathbb{R}_+$ 
with  the  uncertainty relation:
\begin{eqnarray}\label{sosb}
 \Delta \hat{x}\Delta \hat{p}\geq \frac{1}{2}\Big(1+\alpha (\Delta \hat{x})^2+\beta  (\Delta \hat{p})^2\Big)
\end{eqnarray}
where the parameters $\alpha$ and $\beta$ are real positive numbers. 

The motivations for this study derive from a series of works  devoted to the relation
%Physical relevance of 
 \eqref{sos0}. Indeed,
already in  \cite{Kempf1}  Kempf et {\it al} investigated \eqref{sos0} for the 
particular case $\alpha= 0$ with
\begin{eqnarray}\label{conard}
\Delta \hat{x} \Delta \hat{p} \ge \frac{1}{2} (1+ \beta (\Delta \hat{p})^2) 
\label{ucr}
\end{eqnarray}
and led to the conclusion that the energy levels of a given system can deviate significantly from the 
usual quantum mechanical case once energy scales become comparable to the scale $\sqrt \beta.$
 Although the onset of this scale is an empirical question, it is presumably set by quantum gravitational effects.
	In another work \cite{Kempf2},
Kempf, for the same model with $\alpha= 0$, led to the conclusion  that the anomalies observed with 
fields over unsharp coordinates might be testable if the onset of strong gravity effects is not too far above the currently
 experimentally accessible scale about $10^{-18}m$, rather than at the Planck scale of $10^{-35}m.$
 More recently, in \cite{Kempf4}, it was shown that  similar relation can be applied to discrete models 
of matter or space-time, including loop quantum cosmology. For more motivations, see  \cite{grossetal},
 \cite{grossetal1} and \cite{grossetal2}, but also \cite{Hir}, \cite{Quesne},  \cite{Lay} and \cite{Stetsko} and references therein.

In this work, we investigate how such a deformation may affect  main properties, 
 e.g. energy spectrum, of a simple physical system like a harmonic oscillator.

The paper is organized as follows. In section 2, we introduce
 a reformulation of the $\theta(\hat{x},\hat{p})-$deformation in terms of
a $q-$deformation allowing to easily deduce the energy spectrum of the induced deformed harmonic oscillator. 
Then it is proved that the deformed position and momentum operators admit a one-parameter family of self-adjoint extensions. 
These operators engender new families of deformed Hermite polynomials generalizing usual $q-$ Hermite polynomials.
Section 3 is devoted to the matrix element computation. Finally, in section 4, we provide a $su(2)-$algebra representation 
of the considered
deformation. Section 5 deals with concluding remarks.
%%%%%%%%%%%%%%%%
%%%%%%%%%%%%%%%%%%
%%%%%%%%%%%%%%%%%%
%%%%%%%%%%%%%%%%%%
%%%%%%%%%%%%%%%%%%
%%%%%%%%%%%%%%%%%%
%%%%%%%%%%%%%%%%%%%%%
%%%%%%%%%%%%%%%%%%%%%%%%%%%%%%
\section{ $q$-like realization}
It is worth noticing that such a $\theta(x,p)-$deformation \eqref{sos0} admits an interesting $q-$like realization via the
 following re-parameterization
 of deformed creation and annihilation operators:
% Now consider the harmonic oscillator problem with non canonical commutation relation \eqref{sos0}. We can been defined in this situation
%   the   deformed creation, annihilation operator given respectively by the following relation
\begin{eqnarray}
\hat{b}^\dag=\frac{1}{2}(m_\alpha \hat{x}-im_\beta \hat{p}),\quad \hat{b}=\frac{1}{2}(m_\alpha \hat{x}+im_\beta \hat{p})
\end{eqnarray}
satisfying the peculiar q-Heisenberg commutation 
relation:
\begin{eqnarray}\label{comr}
\hat{b}\hat{b}^\dag-q\hat{b}^\dag\hat{b}=1
\end{eqnarray}
where the parameter $q$ is written in the form
\begin{eqnarray}
q=\frac{1+\sqrt{\alpha\beta}}{1-\sqrt{\alpha\beta}}\geq 1
\end{eqnarray}
and the quantities $m_\alpha$ and $m_\beta$ are given by 
\begin{eqnarray}
m_\alpha=\sqrt{2\alpha\Big(\frac{1}{\sqrt{\alpha\beta}}-1\Big)},\quad m_\beta=\sqrt{2\beta\Big(\frac{1}{\sqrt{\alpha\beta}}-1\Big)}.
\end{eqnarray}
%%%%%%%%%%%%%%%%%%%%%%%%%%%%%%%%%%%%%%%%%%%%%%%%%%%%
With this consideration, the spectrum of the induced harmonic oscillator Hamiltonian $\hat{H}=\hat{b}\hat{b}^\dag+\hat{b}^\dag\hat{b}$ 
 is given by
\begin{eqnarray}\label{folie}
\mathcal{E}_n=\frac{1}{2}([n]_q+[n+1]_q)
\end{eqnarray}
where the $q-$number $[n]_q$ is defined by 
%\begin{eqnarray}
$[n]_q=\frac{1-q^n}{1-q}$.
Let $\mathcal{F}$  be a $q-$ deformed Fock space and $\{|n, q\rangle\;|\;n\in \mathbb{N}\bigcup \{0\}\}$ be its orthonormal  basis. 
The actions of $\hat{b}$, $\hat{b}^\dagger$  on  $\mathcal{F}$ are given by 
\begin{eqnarray}\label{Harmonic2b}
\hat{b}|n,q\rangle= \sqrt{[n]_q}|n-1, q\rangle,\;\mbox{ and }\;\;\hat{b}^\dagger|n, q\rangle=\sqrt{[n+1]_q}|n+1, q\rangle,\; 
\end{eqnarray}
where  $|0, q\rangle$ is a normalized vacuum:
\begin{eqnarray}
 \hat{b}|0, q\rangle= 0,\qquad \langle q, 0|0, q\rangle=1.
\end{eqnarray}
The Hamiltonian operator $\hat{H}$ acts on the states $|n,q\rangle$ to give: $\hat{H}|n,q\rangle= \mathcal{E}_n|n,q\rangle.$ 
%%%%%%%%%%%%%%%%%%%%%%%%%%%%%%%%%%%%%%%%%%%%%%%%%%%%
%%%%%%%%%%%%%%%%%%%%%%%%%%%%%%%%%%%%%%%%%%%%%%%%%%%%
%%%%%%%%%%%%%%%%%%%%%%%%%%%%%%%%%%%%%%%%%%%%%%%%%%%%
%%%%%%%%%%%%%%%%%%%%%%%%%%%%%%%%%%%%%%%%%%%%%%%%%%%%
%%%%%%%%%%%%%%%%%%%%%%%%%%%%%%%%%%%%%%%%%%%%%%%%%%%%
%%%%%%%%%%%%%%%%%%%%%%%%%%%%%%%%%%%%%%%%%%%%%%%%%%%%
\begin{theorem}\label{thm}
 The position operator $\hat{x}$ and momentum operator $\hat{p},$  defined on the Fock space $\mathcal{F},$  are not essentially self-adjoint,
 but have a one-parameter family of self-adjoint extensions. 
\end{theorem}
{\bf Proof:} 
% The matrix elements of  the operators $(\hat{b}+\hat{b}^\dagger)$ and $i(\hat{b}-\hat{b}^\dagger)$ 
% on the basis $|n, q\rangle$ are given by
Consider the following matrix elements of  the position operator $\hat{x}$ and momentum operator $\hat{p}$ :
\begin{eqnarray}
<m,q|\hat{x}|n,q>&:=&<m,q|\frac{1}{m_\alpha}(\hat{b}^\dag+\hat{b})|n,q>\cr
&=&
\frac{1}{m_\alpha}\sqrt{[n+1]_q}\delta_{m,n+1}+\frac{1}{m_\alpha}\sqrt{[n]_q}\delta_{m,n-1}
\end{eqnarray}
\begin{eqnarray}
<m,q|\hat{p}|n,q>&:=&<m,q|\frac{i}{m_\beta}(\hat{b}^\dag-\hat{b})|n,q>\cr
&=&
\frac{i}{m_\beta}\sqrt{[n+1]_q}\delta_{m,n+1}-\frac{i}{m_\beta}\sqrt{[n]_q}\delta_{m,n-1}.
\end{eqnarray}
% \begin{eqnarray}
% %<m,q|(\hat{b}+\hat{b}^\dagger)|n,q>&:=&
% <m,q|\frac{1}{m_\alpha}(\hat{b}^\dag+\hat{b})|n,q>
% &=&
% \frac{1}{m_\alpha}\sqrt{[n+1]_q}\delta_{m,n+1}+\frac{1}{m_\alpha}\sqrt{[n]_q}\delta_{m,n-1}\nonumber\\
% \end{eqnarray}
% \begin{eqnarray}
% %<m,q|i(\hat{b}-\hat{b}^\dagger)|n,q>&:=&
% <m,q|\frac{i}{m_\beta}(\hat{b}^\dag-\hat{b})|n,q>
% &=&
% \frac{i}{m_\beta}\sqrt{[n+1]_q}\delta_{m,n+1}-\frac{i}{m_\beta}\sqrt{[n]_q}\delta_{m,n-1}.\nonumber\\
% \end{eqnarray}
Setting $x_{n,\alpha}=\frac{1}{m_\alpha}\sqrt{[n]_q}$ and $x_{n,\beta}=\frac{1}{m_\beta}\sqrt{[n]_q}$, then
the position operator $\hat{x}$ and momentum operator $\hat{p}$ 
   can be represented by the two following symmetric Jacobi matrices, respectively:
\begin{eqnarray}\label{jacobir}
M_{\hat{x},q,\alpha}= \left(\begin{array}{cccccc}0&x_{1,\alpha}&0&0&0&\cdots\\x_{1,\alpha}&0&x_{2,\alpha}&0&0&\cdots\\0&x_{2,\alpha}&0&x_{3,\alpha}&0&\cdots\\\vdots&\ddots&
\ddots&\ddots&\ddots&\ddots
       \end{array}\right)
\end{eqnarray}
and
\begin{eqnarray}\label{jacobic}
M_{\hat{p},q,\beta}= \left(\begin{array}{cccccc}0&-ix_{1,\beta}&0&0&0&\cdots\\ix_{1,\beta}&0&-ix_{2,\beta}&0&0&\cdots\\0&ix_{2,\beta}&0&-ix_{3,\beta}&0&\cdots\\\vdots&\ddots&\ddots&\ddots&\ddots&\ddots
       \end{array}\right).
\end{eqnarray}
The quantity  $|x_{n,\alpha}|=\frac{1}{m_\alpha}\Big|\frac{1-q^n}{1-q}\Big|^{1/2}$ is not bounded since $q>1$ by definition, and
$
\lim\limits_{n\rightarrow \infty}|x_{n,\alpha}|=\infty.
$
 Considering the series $y_{\alpha}=\sum\limits_{n=0}^\infty1/x_{n,\alpha},$ we get
\begin{eqnarray*}
\overline{\lim_{n\to\infty}}\left(\frac{1/x_{n+1}}{1/x_n}\right)=q^{-1/2}<1
\end{eqnarray*}
and, hence, the series $y_\alpha$ converges. Besides, as  the quantity $q^{-1}+q> 2,$ 
\begin{eqnarray}
0< x_{n+1,\alpha}x_{n-1,\alpha}=\frac{1}{m_\alpha^2(1-q)}\Big[1-q^n(q^{-1}+q)+q^{2n}\Big]^{1/2}
<x_{n,\alpha}^2
\end{eqnarray}
Hence, the Jacobi matrices  in (\ref{jacobir}) and (\ref{jacobic}) are not self-adjoint 
(Theorem 1.5., Chapter VII in Ref. \cite{Berezanskii}).
The deficiency indices of these operators are equal to $(1,1)$.  One concludes that the position operator $\hat{x}$ and the momentum
operator $\hat{p}$ are no longer 
essentially self-adjoint
but have each a one-parameter family of self-adjoint extensions instead. This
means that their deficiency subspaces are one-dimensional.  $\square$

 Besides, in this case, the deficiency subspaces $N_z$,
$Imz \ne 0,$ are defined  by the generalized vectors $|z \rangle=\sum\limits_0^\infty P_n (z)||n,q\rangle$ such that 
 \cite{Berezanskii}, \cite{Burban}:
\begin{eqnarray}\label{pol}
\sqrt{[n]_q} P_{n-1}(z) + \sqrt{[n+1]_q} P_{n+1}(z) = zP_n(z)
\end{eqnarray}
with the initial conditions $P_{-1}(z) = 0,\;\; P_0(z) = 1$.
\begin{itemize}
\item In the position representation,  the states $|x,q>$ such that 
\begin{eqnarray}\label{lab3}
\hat{x}|x,q>=x|x,q>,\mbox{ and}\,\,|x,q>=\sum_{n=0}^\infty P_{n,q}(x)|n,q>,
\end{eqnarray}
transforms the relation (\ref{pol}) into
\begin{eqnarray}
m_\alpha x P_{n,q}(x)&=&\sqrt{[n+1]_q} P_{n+1,q}(x)+\sqrt{[n]_q} P_{n-1,q}(x)\cr
n &=& 0, 1,\ldots;\; P_{-1,q}(x)=0,\;P_{0,q}(x)=1 
\end{eqnarray}
% \subsection{Position representation}
% Consider in this part the states $|x,q>\in\mathcal{H}$ such that 
% \begin{eqnarray}\label{lab3}
% \hat{x}|x,q>=x|x,q>,\mbox{ and}\,\,|x,q>=\sum_{n=0}^\infty P_{n,q}(x)|n,q>.
% \end{eqnarray}
% Using relation \eqref{Lyon} and \eqref{lab3} we arrive to reccurence relation between functions $ P_{n,q}(x)$ as
% \begin{eqnarray}
% m_\alpha x P_{n,q}(x)&=&\sqrt{[n+1]_q} P_{n+1,q}(x)+\sqrt{[n]_q} P_{n-1,q}(x)\cr
% n &=& 0, 1,\ldots;\; P_{-1,q}(x)=0,\;P_{0,q}(x)=1. 
% \end{eqnarray}
giving
\begin{eqnarray}\label{sak}
2\gamma(x,q)P_{n,q}\Big(\frac{2\gamma(x,q)}{(1-q)^{1/2}m_\alpha}\Big)&=&(1-q^{n+1})^{\frac{1}{2}}P_{n+1,q}\Big(\frac{2\gamma(x)}{(1-q)^{1/2}m_\alpha}\Big)\cr
&+&(1-q^{n})^{\frac{1}{2}}P_{n-1,q}\Big(\frac{2\gamma(x,q)}{(1-q)^{1/2}m_\alpha}\Big)
\end{eqnarray}
where $2\gamma(x,q)=(1-q)^{1/2}m_\alpha x.$ 

Setting $\widetilde{P}_{n,q}(\gamma(x,q))=P_{n,q}\Big(\frac{2\gamma(x,q)}{(1-q)^{1/2}m_\alpha}\Big)$, the
equation \eqref{sak} can be re-expressed as
%\end{document}
\begin{eqnarray}\label{sakl}
2\gamma(x,q)\widetilde{P}_{n,q}(\gamma(x,q))&=&(1-q^{n+1})^{\frac{1}{2}}\widetilde{P}_{n+1,q}(\gamma(x,q))\cr
&+&(1-q^{n})^{\frac{1}{2}}\widetilde{P}_{n-1,q}(\gamma(x,q)).
\end{eqnarray}
%\end{document}
Finally, putting $(q;q)_n^{1/2}\widetilde{P}_{n,q}(\gamma(x,q))=H_{n,q}(x),$ the formula (\ref{sakl})
 recalls the  recurrence relation  satisfied  by $q-$Hermite polynomials:
\begin{eqnarray}
2xH_{n,q}(x)=H_{n+1,q}(x)+(1-q^n)H_{n-1,q}(x)
\end{eqnarray}
where $(q;q)_n=(1-q)(1-q^2)\cdots (1-q^n).$
% \begin{proposition}
% The states $(q;q)_n^{-1/2}H_{n,q}(x)$ solve the eigenvalue problem of 
% the harmonic oscillator  in the deformed Heisenberg states given by the commutation relation \eqref{sos0} with spectrum given in equation \eqref{folie}.
% \end{proposition}
\item
In the momentum representation, the state $|p,q>$ such that
\begin{eqnarray}\label{lab5}
\hat{p}|p,q>=p|p,q>,\mbox{ and}\,\,|p,q>=\sum_{n=0}^\infty Q_{n,q}(p)|n,q>
\end{eqnarray}
leads to the following reccurence relation between functions $Q_n(x,q):$ 
\begin{eqnarray}
im_\beta pQ_{n,q}(p)&=&\sqrt{[n+1]_q}Q_{n+1,q}(p)-\sqrt{[n]_q}Q_{n-1,q}(p)\cr
n&=&0, 1,\ldots;\; Q_{-1,q}(p)=0,\;Q_{0,q}(p)=1. 
\end{eqnarray}
This equation can be also re-expressed as
\begin{eqnarray}\label{saki}
\tilde{\gamma}(\tilde{p},q)Q_{n,q}(p)=(1-q^{n+1})^{\frac{1}{2}}Q_{n+1,q}(p)-(1-q^{n})^{\frac{1}{2}}Q_{n-1,q}(p)
\end{eqnarray}
where $\tilde{\gamma}(\tilde{p},q)=(1-q)^{1/2}m_\beta \tilde{p},$ $\tilde{p}=ip.$ 

 Setting $\widetilde{Q}_{n,q}(\tilde{\gamma}(\tilde{p},q))=Q_{n,q}\Big(\frac{2\tilde{\gamma}(\tilde{p},q)}{(1-q)^{1/2}m_\alpha}\Big)$, 
then the
equation \eqref{saki} yields
\begin{eqnarray}
2\tilde{\gamma}(\tilde{p},q)\widetilde{Q}_{n,q}(\tilde{\gamma}(\tilde{p},q))&=&(1-q^{n+1})^{\frac{1}{2}}\widetilde{Q}_{n+1,q}(\tilde{\gamma}(\tilde{p},q))\cr
&&-(1-q^{n})^{\frac{1}{2}}\widetilde{Q}_{n-1,q}(\tilde{\gamma}(\tilde{p},q)).
\end{eqnarray}
Letting $(q;q)_n^{1/2}\widetilde{Q}_{n,q}(\tilde{\gamma}(\tilde{p},q))=H_{n,q}(ip),$ we arrive at the recurrence relation 
 satisfied by the complex $q-$polynomials $H_{n,q}(ip)$ given by
\begin{eqnarray}
2ipH_{n,q}(ip)=H_{n+1,q}(ip)-(1-q^n)H_{n-1,q}(ip).
\end{eqnarray}
\end{itemize}
\begin{remark} The following is worthy of attention:
%We now give the condition that satisfy the operators $\hat{x}$ and $\hat{p}$ regarding the hermiticity condition.
\begin{enumerate}
\item[(i)] In the $x-$space
where the momentum operator is defined  by the relation
 \begin{eqnarray}\label{aa1}
\hat{p}:=-i\theta(\hat{x},\hat{p})\partial_x,
\end{eqnarray} 
 any function $\Psi_q(x)$ in $x-$representation  can be expressed in terms
of its 
 analog $\Psi_q(p)$ in $p-$representation by the relation
\begin{eqnarray}
\Psi_q(x)=\int_{-\infty}^{\infty}dp\, \exp\Big(\frac{ip}{\alpha\sigma(p)}\arctan\frac{x}{\sigma(p)}\Big)\Psi_q(p), 
\end{eqnarray}
where $ \sigma(p)=\sqrt{p^2+\frac{1}{\alpha}}$. Defining the Hilbert space inner product as
\begin{eqnarray}
<f,g>=\int\,\frac{dx}{\theta(\hat{x},\hat{p})} \bar{f}_q(x)g_q(x)
\end{eqnarray}
one can readily prove that $\hat{p}$ reverts the property of a Hermitian operator.
\item[(ii)]
Analogously, in the $p-$space 
\begin{eqnarray}\label{aa2}
\hat{x}:=i\theta(\hat{x},\hat{p})\partial_p
\end{eqnarray}
 and
\begin{eqnarray}
\Psi_q(p)=\int_{-\infty}^{\infty}dx\, \exp\Big(\frac{-ix}{\alpha\sigma(x)}\arctan\frac{p}{\sigma(x)}\Big)\Psi_q(x).
\end{eqnarray}
The appropriate inner product, in the momentum space, rendering
the operator $\hat{x}$ Hermitian is defined as
\begin{eqnarray}
<f,g>=\int\,\frac{dp}{\theta(\hat{x},\hat{p})} \bar{f}_q(p)g_q(p)
\end{eqnarray}
with the condition 
$\lim\limits_{x\rightarrow -\infty}\Psi_q(x)=\lim\limits_{x\rightarrow \infty}\Psi_q(x)=0.$
\end{enumerate}
\end{remark}
\section{Matrix elements} 
From the natural actions of $q-$deformed position operator $\hat{x}$ and momentum operator $\hat{p}$  on
the  basis vectors $|n,q> \in \mathcal{F}:$
\begin{eqnarray}
\hat{x}|n,q>=\frac{1}{m_\alpha}(\hat{b}+\hat{b}^\dag)|n,q>,\quad 
\hat{p}|n,q>=\frac{i}{m_\beta}(\hat{b}^\dag-\hat{b})|n,q>
\end{eqnarray}
we immediately deduce
the matrix elements 
\begin{eqnarray}\label{sol1}
<m,q|\hat{b}^{\dag l}\hat{b}^{r}|n,q>=\sqrt{\frac{\Gamma_q(n+1)\Gamma_q(n-r+l+1)}{\Gamma_q(n-r+1)\Gamma_q(n-r+1)}}\delta_{m,n-r+l}\\
\label{sol2}<m,q|\hat{b}^{r}\hat{b}^{\dag l}|n,q>=\sqrt{\frac{\Gamma_q(n+l+1)\Gamma_q(n+l+1)}{\Gamma_q(n+1)\Gamma_q(n-r+l+1)}}\delta_{m,n-r+l}
\end{eqnarray}
where $\Gamma_q(.)$ is the $q$-Gamma function.  
 Denoting by $:\,:$ the normal ordering, then the expectation value of normal ordering product 
of $\hat{x}^l\hat{p}^r$ can be computed by the following relation:
\begin{eqnarray}
<m,q|:\hat{x}^l\hat{p}^r:|n,q>=\frac{i^r}{m_\alpha^lm_\beta^r}\sum_{s=0}^l\sum_{t=0}^r C_l^s C_r^t<m,q|\hat{b}^{\dag l-s+t}\hat{b}^{s+r-t}|n,q>\cr
\end{eqnarray}
which can be  given explicitly by using relation \eqref{sol1}. Then it becomes a matter of computation to determine 
the basis operators  in terms of $\hat{b}$ and $\hat{b}^\dag$ as follows:
\begin{eqnarray}
|m,q><n,q|=:\frac{\hat{b}^{\dag m}}{\sqrt{[m]_q!}}e^{-\hat{b}^\dag\hat{b}}\frac{\hat{b}^{ n}}{\sqrt{[n]_q!}}:.
\end{eqnarray}
%%%%%%%%%%%%%%%%%%%%%%%%%%%%%%%%%%%%%%%%%%%%%%%%%%%%
%%%%%%%%%%%%%%%%%%%%%%%%%%%%%%%%%%%%%%%%%%%%%%%%%%%%
\section{$su(2)-$algebra representation }
Turning back to the standard expression of  the harmonic oscillator Hamiltonian operator, i.e.
 $\hat{H}=\hat{a}^\dag \hat{a},$ such that
\begin{eqnarray}
\hat{a}=\frac{1}{\sqrt{2}}(\hat{x}+i\hat{p}),\,\,\, \hat{a}^\dag=\frac{1}{\sqrt{2}}(\hat{x}-i\hat{p}),
\end{eqnarray}
we get explicitly
\begin{eqnarray}
\hat{H}=\frac{1}{2}\Big[(1+\alpha)\hat{x}^2+(1+\beta)\hat{p}^2+1\Big]
\end{eqnarray}
giving the simpler form $\hat{H}=\frac{1}{2}$ when $\alpha=-1$ and $\beta=-1.$ 

From \eqref{aa1} and \eqref{aa2},  the Hamiltonian $H$ can be considered as non-local  and we can define
\begin{eqnarray}
\hat{H}^{\mbox{loc}}:=\hat{H}(\theta,\partial_x\theta,\partial_x^2\theta\cdots, x,\partial_x,\partial_x^2,\cdots,\alpha,\beta)
\end{eqnarray}
with 
\begin{eqnarray}
\theta,\partial_x\theta,\cdots=f(\theta,\partial_x\theta,\partial_x^2\theta\cdots, x,\partial_x,\partial_x^2,\cdots,\alpha,\beta).
\end{eqnarray}
adding some nonlinearity to the Hamiltonian operator nonlocality.

%\begin{proposition}
 Assume  the parameters $\alpha$ and $\beta $  satisfy the condition:  $|\alpha|<<1,|\beta|<<1$
 and put  $\tilde{\alpha}=\alpha$ and $\tilde{\beta}=-\beta.$
Then  $\hat{x},\,\hat{p},\,\theta$ can be viewed as the elements of 
$su(2)-$algebra, i.e. 
\begin{eqnarray}
 [\hat{x},\hat{p}]=i\theta,\,\,\,[\hat{p},\theta]=i\alpha\{\hat{x},
\theta\}=i\tilde{\alpha}\hat{x}, \,\,\,[\theta,\hat{x}]=-i\beta\{\hat{p},\theta\}=i\tilde{\beta}\hat{p}.
\end{eqnarray}
%\end{proposition}
Let $\hat{\overrightarrow{J}}:=(\hat{x},\hat{p},\theta)$
% the three component vectors which can be view as the components of
be the angular momentum such that there exist    states $|j,m>\in\mathcal{F}$ satisfying the condition $<j,m|j,m'>=\delta_{mm'}$.
Define the operators $\hat{J}_+$ and $\hat{J}_-$ by
\begin{eqnarray}
\hat{J}_+:=\frac{1}{\beta}\hat{x}+\frac{i}{\alpha}\hat{p},\,\,\,\, \hat{J}_-:=\frac{1}{\beta}\hat{x}-\frac{i}{\alpha}\hat{p}.
\end{eqnarray}
\begin{proposition}
There exists an arbitrary number $\nu$ such that
\begin{eqnarray}\label{ddd}
&\hat{J}_-|j,m>=C_-(m,j)|j,m-\nu>,\,\,\,\,
\hat{J}_+|j,m>=C_+(m,j)|j,m+\nu>,\cr 
&\theta|j,m>=f(m,j)|j,m>
\end{eqnarray}
where $C_-(m,j),$ $C_+(m,j)$ and $f(m,j)$ are three constants depending on $j$ and $m$.
\end{proposition}
The parameters $j$ and $m$ depend  on $\alpha$ and $\beta.$ 
The unitary representation of $su(2)-$ algebra, $\{|j,m>,\,\, j\in\mathbb{N},\,\, -j\leq m\leq j\},$ is infinite dimensional.
% We note that this unitary representation is necessarily infinite dimensional. 
 The operators $\{\hat{x},\hat{p},\theta\}$ act on the Fock space $\mathcal{H}=\{|j,m>/m\in\mathbb{N}\cup\{0\}\}$ 
following \eqref{ddd}. Note that $\theta$ and $\hat{\overrightarrow{J}^2}=(1+2\alpha)\hat{x}^2+(1+2\beta)\hat{p}^2+1$ commute.
Therefore,  $\hat{\overrightarrow{J}^2}$ and $\hat{H}$ commute too. Besides,
the following commutation relations are in order:
\begin{eqnarray}\label{mmm}
[\theta,\hat{J}_+]=\hat{x}+i\hat{p},\,\,\,\,[\theta,\hat{J}_-,]=-(\hat{x}-i\hat{p}).
\end{eqnarray}
%\begin{proposition}
%%%%%%%%%%%%%%%%%%%%%%%%%%%%%%%%%%%%%%%%%%%%%%%%%%%%
In the interesting particular case where  $\alpha=\beta$,  the relations \eqref{mmm} are reduced to
\begin{eqnarray}
[\theta,\hat{J}_+]=\alpha \hat J_+,\,\,\,\,[\theta,\hat{J}_-,]=-\alpha\hat J_-,\,\,\,\, [\hat{J}_+,\hat{J}_-]=2\alpha^{-2}\theta.
\end{eqnarray}
 Taking $f(m,j)=m$ yields the condition
\begin{eqnarray}
\theta\hat{J}_+|j,m>=(m+\alpha)\hat{J}_+|j,m>,\,\,\, \theta\hat{J}_-|j,m>=(m-\alpha)\hat{J}_+|j,m>.
\end{eqnarray}
Besides, we have
\begin{eqnarray}
\hat{J}_+|j,m>=C_+|j,m+\alpha>,\,\,\,\hat{J}_-|j,m>=C_-|j,m-\alpha>
\end{eqnarray}
where
\begin{eqnarray}
C_+=\sqrt{(j-m)(j+m+\alpha)},\quad C_-=\sqrt{(j+m)(j-m+\alpha)}.
\end{eqnarray}
%\end{proposition}
The eigenfunctions of the Hamiltonian $\hat{H}$ in the position and momentum representations are given, respectively, by 
\begin{eqnarray}
\Psi_{j,m}(x)=<x|j,m>,\,\,\,\, \Psi_{j,m}(p)=<p|j,m>
\end{eqnarray}
solution of the equation
\begin{eqnarray}
\hat{H}\Psi_{j,m}(x)=\frac{\alpha^2}{2}\sqrt{(j+m)(j-m)(j+m+\alpha)(j-m+\alpha)}\Psi_{j,m}(x)
\end{eqnarray}
 easily obtainable by solving the eigenvalue problem for the Casimir operator
$\hat{J}_+\hat{J}_-$. 
Furthermore, we get
\begin{eqnarray}
\hat{J}^2\Psi_{j,m}(x)=2\alpha^2\sqrt{(j+m)(j-m)(j+m+2\alpha)(j-m+2\alpha)}\Psi_{j,m}(x).
\end{eqnarray}
%%%%%%%%%%%%%%%%%%%%%%%%%%%%%%%%%%%%%%%%%%%%%%%%%%%%
\section{Concluding remarks}
In work, we have introduced
 a reformulation of the ${\theta}(\hat{x},\hat{p})-$deformation in terms of
a $q-$deformation allowing to easily deduce the energy spectrum of the induced deformed harmonic oscillator. 
Then we have proved  that the deformed position and momentum operators admit a one-parameter family of self-adjoint extensions. 
These operators have engendered new families of deformed Hermite polynomials generalizing usual $q-$ Hermite polynomials.
We have also computed relevant matrix elements. Finally,  a $su(2)-$algebra representation 
of the considered
deformation has been investigated and discussed. 
\subsection*{Acknowledgment}
This work is partially supported by the ICTP through the
OEA-ICMPA-Prj-15. The ICMPA is in partnership with the Daniel
Iagolnitzer Foundation (DIF), France. MNH expresses his gratefulness
to Professor A. Odzijewicz and all his staff for their hospitality and the good organization
of the Workshops in Geometric Methods in Physics.
%%%%%%%%%%%%%%%%%%%%%%%%%%%%%%%%%%%%%%%%%%%%%%%%%%%%


\begin{thebibliography}{1}
%%%%%%%%%%%%%%%%%%%%%%%%%%%%%%%%%%%%%%%%%%%%%%%%%%%%
%%%%%%%%%%%%%%%%%%%%%%%%%%
\bibitem{grossetal1} D. Amati, M. Ciafaloni, G. Veneziano.: (1989), {\em Can spacetime be probed below the string size} 
Phys. lett. {B 216} 41.
%%%%%%%%%%%%%%%%%%%%%%%%%%%%%%%%%%%%%%%%%%%%%%%%%%%%
%%%%%%%%%%%%%%%%%%%%%%%%%%%%%%%%%%%%%%%%%%%%%%%%%%%%%%%%%%%%%%%%%%
\bibitem{Berezanskii} Ju. M. Berezanski\'{i}, {\it Expansions in Eigenfunctions of Selfadjoint Operators}, (Amer. Math. Soc., Providence, Rhode Island, 1968).
%%%%%%%%%%%%%%%%%%%%%%%%%%%%%%%%%%%%%%%%%%%%%%%%%%%%
%%%%%%%%%%%%%%%%%%%%%%%%%%%%%%%%%%%%%%%%%%%%%%%%%%%%%%%%%%%%%%%%%%%%%%%%%%%
\bibitem{Burban} Burban I. M.: {\em Generalized $q-$deformed oscillators, $q-$Hermite polynomials, generalized coherent states}.
%%%%%%%%%%%%%%%%%%%%%%%%%%%%%%%%%%%%%%%%%%%%%%%%%%%%
%%%%%%%%%%%%%%%%%%%%%%%%%%%%%%%%%%%%%%%%%%%%%%%%%%%%%%%%%%%%%%%%%%%%%
%%%%%%%%%%%%%%%%%%%%%%%%%%%%
%%%%%%%%%%%%%%%%%%%%%%%%%%%%%
\bibitem{Doplicher}
 Doplicher, S.  Fredenhagen, K. and   E. Roberts, J.:  (1995),
{\em The Quantum Structure of Spacetime at the Planck
Scale and Quantum Fields},  {\it Comm. Math. Phys.} 172, pp.~187-220.
%%%%%%%%%%%%%%%%%%%%%%%%%%%%%%%%%%%%%%%%%%%%%%%%%%%%%%%%%%%%%%%%%%%%%%%%%%%%%
%%%%%%%%%%%%%%%%%%%%%%%%%%%%%%%%%%%%%%%%%%%%%%%%%%%%%%%%%%%%%%%%%%%%%%%%%%%%%%%%

\bibitem{grossetal2} L. J. Garay.: (1995) {\em Models of neutrino masses and mixings, } Int. J. Mod.  Phys.  {\bf A10}, 145 
%%
%%%%%%%%%%%%%%%%%%%%%%%%%%%%%%%%%%%%%%%%%%%%%%%%%%%%%%%%%%%%%%%%%%%%%%%%%
\bibitem{grossetal} 
D. J. Gross, P. F. Mende.:  (1988){\em The Minimal Length in String Theory,}
Nucl. Phys. {B303}, 407.
 

%%%%%%%%%%%%%%%%%%%%%%%%%%%%%%%%%%%%%%%%%%%%%%%%%%%%%%%%%%%%%%%%%%%%%%
%%%%%%%%%%%%%%%%%%%%%%%%%%%%%%%%%%%%%%%%%%%%%%%%%%%%%%%%%%%%%%%%%%%%%%%%%
%%%%%%%%%%%%%%%%%%%%%%%%%%%%%%%%%%%%%%%%%%%%%%%%%%%%%%%%%%%%%%%%%%%%%%%%%%%%%%
\bibitem{Hir}
Hirshfeld, A.C.  and Henselder, P.:  (2002), {\em Deformation quantization in the teaching of
quantum mechanics},  American Journal of Physics,  70  (5) pp.~537–547, May.
%%%%%%%%%%%%%%%%%%%%%%%%%%%%%%%%%%%%%%%%%%%%%%%%%%%%%%%%%%%%%%%%%%%%%%%%%%%
%%%%%%%%%%%%%%%%%%%%%%
\bibitem{Kempf2}
Kempf, A.: (2000),  {\em A Generalized Shannon Sampling Theorem,
Fields at the Planck Scale as Bandlimited Signals},  Phys. Rev. Lett. 85, pp.~2873  {\it [e-print hep-th/9905114]}.
%%%%%%%%%%%%%%%%%%%%%%%%%%%%%%%
\bibitem{Kempf3}
Kempf, A.: (1998),  {\em On the Structure of Space-Time at the Planck Scale
},  {\it [e-print hep-th/9810215]}.
%%%%%%%%%%%%%%%%%
%%%%%%%%%%%%%%%%%%%%%%%%%%%%%%%%%%%%%%%%%%%%%%%%%%%%%%%%%%%%%%%%%%%%%%%%%%%%
\bibitem{Kempf4}
Kempf, A.: (2011), {\em Generalized uncertainty principles and localization in discrete space},
{\it [e-print hep-th/1112.0994]}.
%%%%%%%%%%%%%%%%%%%%%%%%%%%%
\bibitem{Kempf1}
 Kempf, A. Mangano, G. and  Mann, R.B.: (1995),
 {\em Hilbert Space Representation of the Minimal
Length Uncertainty Relation}, J. Phys. D. 52,  pp.~1108.
%%%%%%%%%%%%
\bibitem{Lay} Lay Nam Chang, Dlordje Minic, Naotoshi Okamura and Tatsu Takeuchi : {\em Exact solutions of the harmonic
oscillator in arbitrary dimensions with minimal length uncertainty relations}, [arXiv:hep-th/0111181].
%%%%%%%%%%%%%%%%%%%%%%%%%%%%
%\bibitem{lebedev}
%Lebedev, N. N.: (1972), {\em Special Functions and Their Application}, Dover Publications, Inc. 180 Varick Street New
%York, N.Y. 10014. Translated and Edited by Silverman, A. R..
%%%%%%%%%%%%%%%%%%%%%%%%%%%%%%%%%%%%%%%%%%%%%%%%%%%%%%%%%%%%%%%%%%%%
%\bibitem{lukierski}
%Lukierski, J., Nowicki, A. and  Ruegg, H.: (1992), {\em Deformed Kinematics and Addition Law For Deformed Velocities}, 
% Phys. Lett. B. 293, pp.~344--352.
%%%%%%%%%%%%%%%%%%%%%%%%%%%%%%%%%%%%%%%%%%%%%%%%%%%%%%%%%%%%%%%%%%%%%%%%%%
%%%%%%%%%%%%%%%%%%%%%%%%%%%%%%%%%%%%%%%%%%%%%%%%%%%%%%%%%%%%%%%%%%%%%
%\bibitem{Moyal}
%Moyal, J. E.: (1949),  {\em  Quantum mechanics as a statistical theory}, 
 %Proc. Cambridge Phil. Soc.  45, 99.
%%%%%%%%%%%%%%%%%%%%%%%%%%%%%%%%%%%%%%%%%%%%%%%%%%%%%%%%%%%%%%%%%%%%%%%%%%%%%%%%%%%
%%%%%%%%%%%%%%%%%%%%%%%%%%%%%%%%%%%%%%%%%%%%%%%%%%%%
%\bibitem{nakahara} 
%Nakahara, M.: (2003), {\em Geometry, Topology and Physics},
%(Institute of Physics Publishing, Bristol and Philadelphia, Second Edition).
%%%%%%%%%%%%%%%%%%%%%%%%%%%%%%%%%%%%%%%%%%%%%%
%%%%%%%%%%%%%%%%%%%%%%%%%%%%%%%%%%%%%%%%%%%
\bibitem{Quesne} Quesne C. and Tkachuk V. M.: {\em Lorentz-covariant deformed algebra
 with minimal length}, [arXiv:hep-th/0612093].
%\bibitem{Witten}
%Seiberg, N.  and Witten, E.: (1999), {\em String theory and noncommutative geometry},
%JHEP 9909, 032  [arXiv:hep-th/9908.142].
%%%%%%%%%%%%%%%%%%%%%%%%%%%%%%%%%%%%%%%%%%%%%%%%%%%%%%%%%
\bibitem{Stetsko} Sketesko M. M and Tkachuk V. M.: {\em Perturbative hydrogen-atom spectrum in deformed
space with minimal
length}, [arXiv:hep-th/0603042].
%%%%%%%%%%%%%%%%%%%%%%%%%%%%%%%%%%%%%%%%%%%%%%%%%%%%
%%%%%%%%%%%%%%%%%
%%%%%%%%%%%%%%%%%%%%%%%%%%%%%%%%%%%%%%%%%%%%%%%%%%%%
%
%%%%%%%%%%%%%%%%%%%%%%%%%%%%%%%%%%%%%%%%%%%%%%%%%%%%
%%%%%%%%%%%%%%%%%%%%%%%%%%%%%%%%%%%%%%%%%%%%%%%%%%%%
%%%%%%%%%%%%%%%%%%%%%%%%%
%%%%%%%%%%%%%%%%%%%%%%%%%%%%%%%%%%%%%%
%%%%%%%%%%%%%%%%%%%%%%%%%%%%%%%%%%%%%%%%%%%%%%%%%%%%
%%%%%%%%%%%%%%%%%%%%%%%%%%%%%%%%%%%%%%%
%%%%%%%%%%%%%%%%%%%%%%%%%%%%%%%%%%%%%%%%%%%%%%%%%%%%
%%%%%%%%%%%%%%%%%%%%%%%%%%%%%%%%%%%%%%%
\end{thebibliography}
\end{document}